\documentclass[english,12pt]{article}
\usepackage{color}
\usepackage{amssymb}
\usepackage{amsmath}
\usepackage{babel}
\usepackage{pifont}
\usepackage
[dvips,
colorlinks=true,
linkcolor=webgreen,%defined below
filecolor=webbrown,%defined below
citecolor=webgreen,%defined below
bookmarksopen=false,
]{hyperref}
\definecolor{webgreen}{rgb}{0,.5,0}
\definecolor{webbrown}{rgb}{.6,0,0}
\usepackage[T1]{fontenc}
\usepackage[cp1250]{inputenc}
\usepackage{array}
\usepackage[dvips]{graphicx}
\usepackage{amssymb}
\date{}
\definecolor{arcolor}{cmyk}{0.05,0.95,0.9,0.1}
\title{An invitation to Quantum Game Theory }
\author{Edward W. Piotrowski\\ Institute of Theoretical Physics,
University of Bia\l ystok,\\ Lipowa 41, Pl 15424 Bia\l ystok,
Poland\\ e-mail: \href{mailto:ep@alpha.uwb.edu.pl}{ep@alpha.uwb.edu.pl}\\
 Jan S\l adkowski \\ Institute of Physics, University of Silesia, \\ Uniwersytecka
4, Pl 40007 Katowice, Poland \\ e-mail:
\href{mailto:sladk@us.edu.pl}{sladk@us.edu.pl} }
\begin{document}
\baselineskip8mm \maketitle
\def\Z{{\bf Z\!\!Z}}
\def\R{{\bf I\!R}}
\def\N{{\bf I\!N}}
\def\C{{\bf I\!\!\!\! C}}

\begin{abstract}
\baselineskip7mm 
Recent development in quantum computation and
quantum information theory  allows to extend the scope of game
theory for the quantum world. The paper presents the history,
basic ideas and recent development in quantum game theory. In this
context, a new application of the Ising chain model is proposed.

\end{abstract}


 \vspace{5mm}\newpage
{\small \begin{flushleft}
 {\bf motto}\\
The man was very appreciative but curious. He asked the farmer why
he called his horse by the wrong name three times.\\
The farmer said, "Oh, my horse is blind, and if he thought he was
the only one pulling he wouldn't even try".
\end{flushleft}
}

\section{Introduction}

Attention to the very physical aspects of information
characterizes  the recent research in quantum computation, quantum
cryptography and quantum communication.  In most of the cases
quantum description of the system provides advantages over the
classical situation. For example, Simon's quantum algorithm to
identify the period of a function chosen by an oracle is more
efficient than any deterministic or probabilistic algorithm
(Simon, 1994), Shor's polynomial time quantum algorithm for
factoring (Shor, 1994) and the quantum protocols for key
distribution devised by Wiener, Bennett and Brassard, and Ekert
are qualitatively more secure against eavesdropping than any
classical cryptographic system (Bennett, and Brassard G., 1984;
Ekert, 1991).

Game theory, the study of (rational) decision making in conflict
situation, seems to ask for a quantum version. For example, games
against nature (Milnor, 1954) should include those for which
nature is quantum mechanical. Does quantum theory present more
subtle ways of playing games? Classical strategies can be pure or
mixed: why cannot they be entangled? Can quantum strategies  be
more successful than classical ones? And if the answer is yes are
they of any practical value? Finally, von~Neumann is one of the
founders of both modern game theory (von Neumann and Morgenstern,
1953) and quantum theory, is that a meaningful coincidence?

\section{Star Trek: The Gambling Episode}

\begin{center}
{ Based on a novel by  David A. Meyer (Meyer, 1999)}
\end{center}

Captain Picard and Q are characters in the popular TV series {\it
Star Trek: The Next Generation\/}. Suppose they play the {\it
spin-flip game} that is a modern version of the penny flip game
(there should be no coins on a starship). Picard is to set an
electron in the spin up state, whereupon they will take turns (Q,
then Picard, then Q) flipping the spin or not, without being able
to see it. Q wins if the spin is  up when they measure the
electron's state.

This is a two-person zero-sum strategic game which might be
analyzed  using the payoff matrix:
$$
\vbox{\offinterlineskip
\halign{&\vrule#&\strut\enspace\hfil#\enspace\cr
\omit&\omit&\omit&$NN$&\omit&$NF$&\omit&$FN$&\omit&$FF$&\omit\cr
\omit&\omit&\multispan9\hrulefill\cr
\omit&\omit&height2pt&\omit&&\omit&&\omit&&\omit&\cr
\omit&$N$&&$-1$&&1&&1&&$-1$&\cr
\omit&\omit&height2pt&\omit&&\omit&&\omit&&\omit&\cr
\omit&\omit&\multispan9\hrulefill\cr
\omit&\omit&height2pt&\omit&&\omit&&\omit&&\omit&\cr
\omit&$F$&&1&&$-1$&&$-1$&&1&\cr
\omit&\omit&height2pt&\omit&&\omit&&\omit&&\omit&\cr
\omit&\omit&\multispan9\hrulefill\cr }}
$$
where the rows and columns are labelled by Picard's and Q's {\it
pure strategies}, respectively; $F$ denotes a flip and $N$ denotes
no flip; and the numbers in the matrix are Picard's payoffs:  1
indicating a win and $-1$ a loss. Q's payoffs can be obtained by
reversing the signs in the above matrix (this is a general feature
of a {\it zero sum game}).

Example:  Q's strategy is to flip the spin on his first turn and
then not flip it on his second, while Picard's strategy is to not
flip the spin on his turn.  The result is that the state of the
spin is, successively: $U$, $D$, $D$, $D$, so Picard wins.

 It is natural to define a two dimensional vector space $V$
with basis $(U,D)$ and to represent player strategies by sequences
of $2\times 2$ matrices.  That is, the matrices
$$
F := \bordermatrix{
                       &\scriptstyle{U} & \scriptstyle{D}          \cr
       \scriptstyle{U} &       0        &        1                 \cr
       \scriptstyle{D} &       1        &        0                 \cr
                  }
\qquad\hbox{and}\qquad N := \bordermatrix{
                       &\scriptstyle{U} & \scriptstyle{D}          \cr
       \scriptstyle{U} &       1        &        0                 \cr
       \scriptstyle{D} &       0        &        1                 \cr
                  }
$$
correspond to flipping and not flipping the spin, respectively,
since we define them to act by left multiplication on the vector
representing the state of the spin. A general {\it mixed strategy}
consists in a  linear combination of $F$ and $N$, which acts as a
$2\times 2$  matrix:
$$
\bordermatrix{
                       &\scriptstyle{U} & \scriptstyle{D}          \cr
       \scriptstyle{U} &      1-p       &        p                 \cr
       \scriptstyle{D} &       p        &       1-p                \cr
                  }
$$
if the player flips the spin with probability $p \in [0,1]$. A
sequence of mixed actions puts the state of the electron into a
convex linear combination $a U + (1-a) D$, $0 \le a \le 1$, which
means that if the spin is measured the electron will be in the
spin-up state with probability $a$. {Q,  having studied quantum
theory, is utilizing a { \it quantum\/ strategy}, namely a
sequence of unitary, rather than stochastic, matrices. In standard
Dirac notation  the basis of $V$ is written $(|U\rangle,|D\rangle
) $. A {\it pure\/} quantum state for the electron is a linear
combination $a|U\rangle + b|D\rangle$, $a,b \in \C$,
$a\overline{a} + b\overline{b} = 1$, which means that if the spin
is measured, the electron will be in the spin-up state with
probability $a\overline{a}$. Since the electron starts in the
state $|U\rangle$, this is the state of the electron if Q's first
action is the unitary operation
$$
U_1 = U(a,b) := \bordermatrix{
                       &\scriptstyle{U} & \scriptstyle{D}          \cr
       \scriptstyle{U} &       a        &        b                 \cr
       \scriptstyle{D} &  \overline{b}  &  -\overline{a}           \cr
             }.
$$} { Captain Picard is  utilizing a {\it classical
 probabilistic strategy } in which he flips the spin with
probability $p$ (he has preferred drill to studying quantum
theory). After his action the electron is in a { mixed\/} quantum
state, i.e., it is in the pure state $b|U\rangle + a|D\rangle$
with probability $p$ and in the pure state $a|U\rangle +
b|D\rangle$ with probability $1-p$. Mixed states are conveniently
represented as {\it density matrices\/}, elements of $V \otimes
V^{\dagger}$ with trace 1; the diagonal entry $(i,i)$ is the
probability that the system is observed to be in state
$|i\rangle$.  The density matrix for a pure state $|\psi\rangle
\in V$ is the projection matrix $|\psi\rangle\langle\psi|$ and the
density matrix for a mixed state is the corresponding convex
linear combination of pure density matrices. Unitary
transformations act on density matrices by conjugation: the
electron starts in the pure state $\rho_0 = |U\rangle \langle U|$
 and Q's first action puts it into the pure state:
$$
\rho_1 = U_1^{\vphantom\dagger} \rho_0 U_1^{\dagger}
       = \left( \begin{array}{cc} a\overline{a} & a\overline{b} \\
                   b\overline{a} & b\overline{b} \end{array}\right).
$$} Picard's mixed action acts on this density matrix, not as a
stochastic matrix on a probabilistic state, but as a convex linear
combination of unitary (deterministic) transformations:
$$
\rho_2 = p F \rho_1 F^{\dagger} + (1-p) N \rho_1 N^{\dagger}=
$$
$$\left(
        \begin{array}{cc} pb\overline{b} + (1-p)a\overline{a} &
                   pb\overline{a} + (1-p)a\overline{b} \\
                   pa\overline{b} + (1-p)b\overline{a} &
                   pa\overline{a} + (1-p)b\overline{b} \end{array}
                 \right).
$$
For $p = {1\over2}$ the diagonal elements of $\rho_2$ are equal to
${1\over2}$.  If the game were to end here, Picard's strategy
would ensure him an expected payoff of 0, independently of Q's
strategy. In fact, if Q were to employ any strategy for which
$a\overline{a} \ne b\overline{b}$, Picard could obtain an expected
payoff of $|a\overline{a} - b\overline{b}| > 0$ by setting $p =
0,1$ according to whether $b\overline{b} > a\overline{a}$, or the
reverse. Similarly, if Picard were to choose $p \ne {1\over2}$, Q
could obtain an expected payoff of $|2p - 1|$ by setting $a = 1$
or $b = 1$ according to whether $p < {1\over2}$, or the reverse.
Thus the mixed/quantum equilibria for the two-move game are pairs
$\bigl([{1\over2}F + {1\over2}N],[U(a,b)]\bigr)$ for which
$a\overline{a} = {1\over2} = b\overline{b}$ and the outcome is the
same as if both players utilize optimal mixed strategies. {But Q
has another move at his disposal ($U_3$) which again transforms
the state of the electron by conjugation to $\rho_3 =
U_3^{\vphantom\dagger}\rho_2 U_3^{\dagger}$. If Q's strategy
consists of $U_1 = U(1/\sqrt{2},1/\sqrt{2}) = U_3$, his first
action puts the electron into a simultaneous eigenvalue 1
eigenstate of both $F$ and $N$, which is therefore invariant under
 any mixed strategy $pF + (1-p)N$ of Picard; and his second
action inverts his first move to give $\rho_3 = |U\rangle\langle
U|$. That is, with probability 1 the electron spin is up!  Since Q
can do no better than to win with probability 1, this is an
optimal quantum strategy for him. All the pairs $$\bigl([pF +
(1-p)N],
       [U(1/\sqrt{2},1/\sqrt{2}),U(1/\sqrt{2},1/\sqrt{2})]\bigr)$$
are mixed/quantum equilibria, with value $-1$ to Picard; this is
why he loses every game. The end. \section{The moral} The
%moral from
practical lesson that  the above fable teaches is that quantum
theory may offer strategies that at least in some cases bring
advantage over classical strategies. Therefore game theorists
should find answers to the following two questions.
\begin{itemize}
\item Under what conditions some players may be able to take the
advantage of quantum tools?
\item Are there genuine quantum games that have no classical
counterparts or origin?
\end{itemize}
It is not easy to give definite answers at the present stage.
Nevertheless one can present some strong arguments for developing
quantum theory of games. Modern technologies are developed mostly
due to investigation into the quantum nature of matter. This means
that we sooner or later may wind up in captain Picard's position
if we are not on alert. Secondly, quantum phenomena probably play
important role in biological and other complex systems (this point
of view is not commonly accepted) and quantum games may turn out
to be an important tool for the analysis of complex systems. There
are also suggestions that quantum-like description  of market
phenomena may be more accurate than the classical (probabilistic)
one (Waite, 2002). The second question can be answered only after
a thorough investigation. A lot of cryptographic problems can be
reformulated in game-like setting. Therefore quantum information
and quantum cryptography should provide us with a case in point.
It is obvious that some classical games can be implemented in such
a way that the set of possible strategies includes strategies that
certainly deserve the the adjective quantum (Du, J. et al, 2002;
Pietarinen, 2002). This process is often referred to as
quantization of the respective standard game. But this is an abuse
of language: we are in fact defining a new game.
\section{Classical game may involve quantum computation}
Let us consider a game of the type of {\it one against all}
(market). The agent buys and sells the same commodity in a
consecutive way at prices dictated by the market. Let us suppose
that the agent predicts with great probability the changes in
price of the commodity in question. If we denote by $h_m$ the
logarithm of the relative prices $\frac{p_m}{p_{m-1}}$ at the
following quotation times, $m=1,2,\ldots$, then the total profit
(loss) of the agent at the moment $k$ is given by the formula:
\begin{equation}
H(n_1,\ldots,n_k):=-\sum\limits_{m=1}^k h_m n_m\, , \label{forha}
\end{equation}
where the series $(n_m)$ takes the value  $0$ or $1$ if the agent
posses  money or the commodity at the moment $m$, respectively. Of
course the series $(n_m)$ defines the agent's strategy in a unique
way. If take the transaction costs (e.g brokerage) into
consideration the the above formula should be replaced by
\begin{equation}
H(n_1,\ldots,n_k):=-\sum\limits_{m=1}^k \bigl(h_m n_m-j\,(
n_{m-1}\oplus n_m)\bigr)\,, \label{forha1}
\end{equation}
where $\oplus$ denotes the addition  modulo 2, $n_0:=0$, and the
constant $j$ is equal to the logarithm of percent cost of the
transaction. An attentive reader certainly notices that
$(\ref{forha1})$ is the hamiltonian of an Ising chain (Feynmann,
1972) (the shift  by the constant $-\frac{1}{2}$ can be absorbed
into the value of $h_m$ and therefore changes the whole formula by
an unimportant constant). Classes of portfolios that correspond to
the strategy $\text{e}^{-\beta H(n_1,\ldots,n_k)}$, that is to the
canonical distribution, were analyzed in (Piotrowski S\l adkowski,
2001a). To determine the profits and correlation of agent's
behavior we have to know the corresponding statistical sum, that
is the logarithm of the product of the transfer matrix $M(m)$:
$$
 \sum\limits_{n_1,\ldots,n_k=0}^{1}M(1)_{0,n_1}M(2)_{n_1,n_2}\cdots
 M(k)_{n_{k-1},n_k}\,,
$$
where $$ M(m)_{n_{m-1},n_{m}}:=\text{e}^{\beta(h_m n_m-j\,
(n_{m-1}\oplus n_m))}\,.
$$
Unfortunately, the matrix $M(m)$ depend on the parameter $m$
(time) through $h_m$ and the solution to proper value problem does
not lead to a compact form of the statistical sum. It is possible
to find the agent's best strategy (that is the ground state of the
hamiltonian) in the limit $\beta^{-1} \longrightarrow 0^+$. Then
the transfer matrix algebra reduces to the (min,+) algebra
(Gaubert and Plus, 1997). Let us call a {\it  potential ground
state} of the Ising chain for a finite $(h_1,\ldots,h_k)$ a
strategy that if supplemented with elements corresponding to
following moments, $k'>k$, can turn out to be the actual ground
state of  of the hamiltonian $H(n_1,\ldots,n_k,\ldots,n_{k'})$.
These states are of the form
$$
(0,1,1,0,1,0,0,n_{k-l+1},n_{k-l+2},\ldots,n_k)
$$
and consist of two parts. The first one is determined by the
series $(h_1,\ldots,h_k)$ and the second of length $l$,
$(n_{k-l+2},\ldots,n_k)$, that can be called the coherence depth
(c.f. the many world interpretation of quantum theory). The later
can be determined only if we know $h_m$ for $m\negthinspace
>\negthinspace k$. Any potential ground state forms an optimal
strategy for the agent that knows only the data up to the moment
$k$. In this case when the transaction cost are non-zero we
"discover" an obvious arbitrage risk, that for example may results
from the finite maturity time of the contracts. Although the above
model is classical it intrisically connected  with quantum
computation (and games) because all calculations for an arbitrage
with non-zero transaction cost should take account of all
potential ground states, number of which  grows exponentially with
the coherence depth. Therefore only quantum computation exploring,
for example,  quantum states (strategies) of the form $$
|\psi\rangle:=\sum\limits_{n_1\ldots n_k=0}^{1}c_{n_1\ldots
n_k}|n_1\rangle\cdots|n_k\rangle
$$ gives hope for an effective  practical implementation of the
strategy. This is an interesting area for further research.

\section{Quantum game theory} Any quantum
system which can be manipulated  by two parties or more and where
the utility of the moves can be reasonably quantified, may be
conceived as a quantum game. A {\it two-player quantum game}\/
$\Gamma=({\cal H},\rho,S_A,S_B,P_A,P_B)$ is completely specified
by the underlying Hilbert space ${\cal H}$ of the physical system,
the initial state $\rho\in {\cal S}({\cal H})$, where ${\cal
S}({\cal H})$ is the associated state space, the sets $S_A$ and
$S_B$ of permissible quantum operations of the two players, and
the { pay-off (utility) functions}\/ $P_A$ and $P_B$, which
specify the pay-off for each player. \\
A {\it  quantum strategy}\/ $s_A\in S_A$, $s_B\in S_B$ is a
quantum operation, that is, a completely positive trace-preserving
map mapping the state space on itself. The quantum game's
definition may also include certain additional rules, such as the
order of the implementation of the respective quantum strategies.
We also exclude the alteration of the pay-off during the game. The
generalization for the N players case is
obvious.\\
 Schematically we have:
$$
\rho \mapsto (s_{A},s_{B}) \mapsto  \sigma \Rightarrow (P_{A},
P_{B})$$

The following concepts  will be used in the remainder of this
lecture. These definitions are fully analogous to the
corresponding definitions in standard game theory. A quantum
strategy  $s_A$ is called {\it dominant strategy}\/ of Alice if
\begin{eqnarray}
        P_A(s_A,s_B')
        &\geq&
        P_A(s_A',s_B')
\end{eqnarray}
for all $s_A'\in S_A$, $s_B'\in S_B$. Analogously we can define a
dominant strategy for Bob. A pair $(s_A,s_B)$ is said to be an
{\it equilibrium in dominant strategies}\/ if $s_A$ and $s_B$ are
the players' respective dominant strategies. A combination of
strategies $(s_A,s_B)$ is called a {\it Nash equilibrium}\/ if
%\begin{mathletters}
\begin{eqnarray}
        P_A(s_A,s_B)&\geq& P_A(s_A',s_B),\\
        P_B(s_A,s_B)&\geq& P_B(s_A,s_B') .
\end{eqnarray}
%\end{mathletters} for all $s_A'\in S_A$, $s_B'\in S_B$.
A pair of strategies $(s_A, s_B)$ is called {\it Pareto
optimal}\/, if it is not possible to increase one player's pay-off
without lessening the pay-off of the other player. A solution in
dominant strategies is the strongest solution concept for a
non-zero sum game. In the Prisoner's Dilemma

$$ \begin{array}{c|cc}
     & \mbox{Bob}: C & \mbox{Bob}: D \\
    \hline
    \mbox{Alice}: C & (3,3) & (0,5) \\
    \mbox{Alice}: D & (5,0) & (1,1)
  \end{array}
$$
(the numbers in parentheses represent the row (Alice) and column
(Bob) player's payoffs, respectively). Defection is the dominant
strategy, as it is favorable regardless what strategy the other
party chooses.

In general the optimal strategy depends on the strategy chosen by
the other party. A Nash equilibrium implies that neither player
has a motivation to unilaterally alter his/her strategy from this
equilibrium solution, as this action will lessen his/her pay-off.
Given that the other player will stick to the strategy
corresponding to the equilibrium, the best result is achieved by
also playing the equilibrium solution. The concept of Nash
equilibria is therefore of paramount importance to studies of
non-zero-sum games. It is, however, only an acceptable solution
concept if the Nash equilibrium is not unique. For games with
multiple equilibria we have to find a way to eliminate all but one
of the Nash equilibria. A Nash equilibrium is not necessarily
efficient. We say that an equilibrium is  Pareto optimal if there
is no other outcome which would make both players better off. Up
to know several dozens of papers on quantum games have been
published. We would like to mention the following problems (lack
of time):
\begin {itemize}
\item The prescription for quantization of games provided by Eisert
and coworkers (Eisert, Wilkens and  Lewenstein, 1999)
is a general one that can be applied to any $2\times2$
game, with the generalization to $2\times n$ games.  (SU(n)
operators are used to represent the players' actions).
\item Quantum theory of information is certainly a serious challenge
to the standard game theory  (eg quantum eavesdropping, quantum
coin tossing).
\item Evolutionary stable strategies. Iqball and Toor analyzed
several important issues that hint that some biological systems
may in fact behave in quantum-like way (Iqbal, and Toor, 2001).
\item Quantum game theory may help solving some hard philosophical paradoxes,
 c.f. the quantum solution to the Newcomb's paradox (Piotrowski and  S\l adkowski, 2002c).
\item The Monty Hall Problem. This is an interesting game based on
a popular TV quiz. The analysis shows that that quantization of a
classical game may be non-unique (Flitney and  Abbott, 2002;
D'Ariano et al, 2002).
\item In the classical Battle of Sexes  Game there is no satisfactory
resolution. In the quantum version the dead-lock may be broken
(Du, J. et al, 2001). Unfortunately we see no way of using it to
solve marriage problems.
\item There are games in which the agents' strategies do not have
adequate descriptions in terms of some Boolean algebra of logic
and theory of probability. They can be analyzed according to the
rules of quantum theory and the result are promising, see e.g. the
Wise Alice game proposed in (Grib and Parfionov, 2002a and 2002b).
Note that this game is a simplified version of the Quantum
Barganing Game (Piotrowski and S\l adkowski, 2002a) restricted to
the "quantum board" of the form $[buy,sell]\times [bid, accept]$.
\item Proposals for using quantum games in market and stock exchange
description (quantum market games)  have already been put forward
(Piotrowski and S\l adkowski, 2001b; 2002a; 2002b). They seem to
be very promising. At present stage, quantum auction  presents a
feasible idea if we neglect costs of implementation.
\item Parrondo's Paradox consists in asymmetrical combination of
doomed games (strategies) so that the resulting new game is not
biased or there even is a winning strategy. It can be used to
increase reliability and stability of electrical circuits and so
on. Quantum Parrondo Games are also interesting (Flitney, Ng and
Abbott, 2002).
\item Quantum gambling. At the present stage of development it
already is feasible to open "quantum casinos" (Goldenberg, Vaidman
and  Wiesner, 1999; Hwang, Ahn, and Hwang, 2001). Quantum gambling
is closely related to quantum logic and can be used to define a
Bayesian theory of quantum probability (Pitowsky, 2002).
\item To our knowledge,  algorithmic combinatorial games,
except for cellular automata,  have been completely ignored by
quantum physicists. This is astonishing because at least some of
the important intractable problems might be attacked and solved on
a quantum computer (even such a simple one player game as
Minesweeper in NP-complete).
\item MUCH MORE to be find at e.g. the Los Alamos preprint data base.
\end{itemize}

\section{Summary and outlook}
We have given examples of interesting possibilities offered by
quantum strategies. In general quantum extension  of a standard
(classical) game is not unique. Most of the published analysis
explore trace preserving completely positive maps as admissible
quantum operations (tactics or strategies). This restriction is
conventional but not necessary. The effect noise and decoherence,
the use of ancillas and algorithmic aspects in quantum games are
the most important areas that invite further research. Quantum
game theory should turn out to be an important theoretical tool
for investigation of various problems in quantum cryptography and
computation, economics or game theory even  if never implemented
in real world.
 Let us quote the Editor's Note to Complexity Digest 2001.27(4)
(http://www.comdig.org):\\
 "It might be that while observing the
due ceremonial of everyday market transaction we are in fact
observing capital flows resulting from quantum games eluding
classical description. If human decisions can be traced to
microscopic quantum events one would expect that nature would have
taken advantage of quantum computation in evolving complex brains.
In that sense one could indeed say that quantum computers are
playing their market games according to quantum rules".

\begin{flushleft}
{\bf REFERENCES}
\end{flushleft}
%\begin{thebibliography}{99}
\begin{description}
\item Bennett, C. H. and Brassard G., (1984). Quantum cryptography:
Public-key distribution and coin tossing, in {\it Proceedings of
the IEEE International Conference on Computers, Systems and Signal
Processing, Bangalore, India, December 1984}, IEEE, New York, p.
175.
\item  D'Ariano, G.M. et al, (2002). The quantum Monty
Hall Problem, {\it Quantum Information and Computing} {\bf 2} 355.
\item  Du, J. et al (2002). Experimental realization of quantum games on a quantum computer
{\it Physical Review Letters} {\bf 88} 137902.
\item Du, J. et al, (2001). Remarks on Quantum Battle of Sexes Game, preprint quant-ph/0103004.
\item  Eisert, J.,  Wilkens, M. and  Lewenstein, M.
(1999). Quantum Games and Quantum Strategies, {\it  Physical
Review Letters }{\bf 83}  3077.
\item  Ekert, A. (1991). Quantum cryptography based on Bell's theorem", {\it  Physical
Review Letters } {\bf 67}  661.
\item Feynmann, R.P. (1972). {\it Statistical physics. A Set
of lectures}, Benjamin Inc., Menlo Park.
\item Flitney, A.P., and  Abbott, D. (2002).
Quantum Version of the Monty Hall Problem, {\it  Physical Review
A} {\bf  65} 062318.
\item Flitney, A.P., and  Abbott, D. (2002). Quantum version of the
Monty Hall problem, {\it  Physical Review A}  {\bf  65}  art. no.
062318 (2002).
\item Flitney, A.P.,  Ng, J., and  Abbott, D. (2002).
 Quantum Parrondo's Games, {\it Physica A} {\bf 314} 384.
\item  Gaubert, S. and Plus, M. (1997). Methods and applications of max-plus linear
algebra, in {\it Lecture Notes in Computer Sciences} no {\bf
1200}, Springer Verlag, New York.
\item  Goldenberg, L., Vaidman, L., and  Wiesner,  S. (1999). Quantum
Gambling, {\it  Physical Review Letters } {\bf 82}  3356.
\item  Grib, A., and  Parfionov, G. (2002a). Can the game be quantum?, preprint
quant-ph/0206178.
\item  Grib, A., and  Parfionov, G. (2002b). Macroscopic quantum
game, preprint quant-ph/0211068.
\item  Hwang, W.Y., Ahn, D., and Hwang, S.W. (2001). Quantum Gambling Using Two Nonorthogonal
States, {\it  Physical Review A} { \bf 64}, 064302.
\item  Iqbal, A., and Toor, A. H. (2001). Evolutionary
stable strategies in quantum games, {\it Physics Letters A} {\bf
280} 249.
\item Meyer, D. (1999). Quantum strategies, {\it  Physical
Review Letters }{\bf 82}  1052.
\item  Milnor, J. (1954). Games against nature, in R. M. Thrall, C. H. Coombs and R. L.
Davis, eds., {\it Decision Processes\/}, John Wiley \& Sons  New
York p. 49.
\item  von Neumann, J. and Morgenstern, O. (1953).
{\it Theory of Games and Economic Behavior}, Princeton University
Press, Princeton .
\item  Pietarinen, A. (2002). Quantum logic and quantum theory in a
game-theoretic perspective, {\it Open Systems and Information
Dynamics }{\bf 9}  273.
\item  Piotrowski,  E.W., and  S\l adkowski, J. (2001a). The
Thermodynamics of Portfolios, {\it Acta Physica Polonica} {\bf
B32} 597.
\item  Piotrowski,  E.W., and  S\l adkowski, J. (2001b).
 Quantum-like approach to financial risk: quantum anthropic
 principle, {\it Acta Physica Polonica} {\bf B32}  3873.
\item  Piotrowski,  E.W., and  S\l adkowski, J. (2002a).
 Quantum Bargaining Games, {\it Physica }{\bf A 308}  391-401.
\item  Piotrowski,  E.W., and  S\l adkowski, J. (2002b).
Quantum Market Games,  {\it Physica }{\bf  A }312 208.
\item  Piotrowski,  E.W. and  S\l adkowski, J. (2002c). Quantum solution to the
Newcomb's paradox, submitttd to Int. J. Game Theor.; preprint
quant-ph/0202074.
\item  Pitowsky, I. (2002). Betting on the Outcomes of Measurements: A Bayesian Theory of Quantum
Probability, preprint quant-ph/0208121.
\item  Simon, D. R. (1994). On the power of quantum computation, in  {\it
Proceedings of the 35th Symposium on Foundations of Computer
Science, Santa Fe}, S. Goldwasser (ed.),  IEEE Computer Society
Press, Los Alamitos, p. 116.
\item  Shor, P. W. (1994). Algorithms for quantum computation:  discrete logarithms and
factoring, in  {\it Proceedings of the 35th Symposium on
Foundations of Computer Science, Santa Fe}, S. Goldwasser (ed.),
IEEE Computer Society Press, Los Alamitos, p. 124.
\item  Waite, S. (2002). {\it "Quantum investing"}, Texere Publishing, London.
\end{description}
%\bibitem[8]{8} E. Farhi and S. Gutmann,
%{\it "Quantum computation and decision trees"}, MIT preprint
%CTP-2651 (1997), quant-ph/9706062.
%\end{thebibliography}
%\vspace*{1cm}

\end{document}